# Light and thermodynamics:

# The three-level laser as an endoreversible heat engine

Peter Muys

*Laser Power Optics, Ter Rivierenlaan 1 P.O. box 4, 2100 Antwerpen Belgium*
peter.muys@gmail.com
+32 477 451916

Abstract:
In the past, a number of heat engine models have been devised to apply the principles of thermodynamics to a laser. The best one known is the model using a negative temperature to describe population inversion. In this paper, we present a new temperature scale not based on reservoir temperatures. This is realized by revealing a formal mathematical similarity between the expressions for the optimum power generated by an endoreversible heat engine and the optimum output-coupled power from a three-level laser resonator. As a consequence, the theory of endoreversibility can be applied to enable a new efficiency analysis of the cooling of high- power lasers. We extend the endoreversibility concept also to the four-level laser model.

1. Introduction

It was found very early in the history of the laser that a three-level optical amplifier could be considered as a heat engine [1] . The justification was based on a quantum model of the lasing element, and hence was named a "quantum heat engine". The engine is working between two reservoirs, the hot and the cold one, at respective (absolute) temperatures $T_H$ and $T_C$. In the ideal case of reversible operation, this engine follows the Carnot law: its efficiency with which heat (pump power) is converted to work (laser power) is limited by

$$\eta = 1 - \frac{T_C}{T_H} \qquad (1)$$

A number of conceptual difficulties were pointed out later [2] , such as the debatable assignment of negative temperatures to the different atomic transitions, since these are out of equilibrium, and hence not suited to the concept of temperature which presupposes thermal equilibrium. So in 2005, it was concluded [2] that the three-level laser was not accessible to thermodynamical modeling as a heat engine.

In 1975, Kafri and Lavine formulate the statement that at threshold, the laser is working with Carnot efficiency, but also simultaneously with zero power [6]. They still used the concept of negative temperatures to describe the population inversion, plus the assumption that the laser was pumped by a (filtered) blackbody source having temperature $T_H$. The ambient temperature was $T_C$.

Around 1980, a new branch of thermodynamics emerged, nowadays called finite-time thermodynamics [3]. This theory takes irreversible parts of a machine explicitly into account. It conceptually splits the machine into peripheral parts which show dissipative properties and into a central part which still forms an ideal Carnot engine. The main dissipative source is located in the heat exchanger which is needed to transport the heat from the hot reservoir at temperature $T_H$ to the machine at temperature $T_{iH}$. The temperature difference $T_H$–$T_{iH}$ is necessary to realize a macroscopic heat flow from the hot source to the machine. And during transport of heat, entropy is inevitably generated. The denomination "finite time" refers to the fact that reversible processes need to progress infinitely slow, in order not to break the thermal equilibrium. In practice however, the energy needs to be transferred in a finite time, which consequentially generates entropy. The parameter $T_{iH}$ became known as the "control parameter" of the machine. The finite-time analysis resulted in lower but more accurate estimates of the machine efficiency. A machine with localized reversible and irreversible parts is called an endoreversible machine. One is able to come up with expressions for the maximum power which such a machine is capable to deliver. The efficiency *at maximum power output* of the engine now turned out to be given by

$$\eta = 1 - \sqrt{\frac{T_C}{T_H}} \tag{2}$$

instead of the Carnot value (1) which represents the overall maximum efficiency of the engine. It is a central observation in finite-time thermodynamics that at maximum efficiency the machine is working completely reversibly and hence infinitely slowly, meaning that , somewhat awkward, at maximum efficiency the power output of the engine is zero. Eq. (2) is known as the Curzon-Ahlborn efficiency.

In 1984, Kosloff takes the next step, and proposes a quantum model of the three-level laser, thereby pointing to its endoreversible character, deducing the maximum machine power in the Curzon-Ahlborn form, and also finding maximum machine efficiency under the Carnot form at the laser threshold. His control parameter is however not linked with the engineering parameters of a laser cavity. He also did not mention the power output of this quantum model. The reason is that he considered only the thermodynamical quantities in terms of quantum observables, so he did not leave the framework of the quantum model. Later work stayed concentrated on the quantummechanical modeling of the irreversibilities .

In this paper, we point out that the three-level laser is an endoreversible engine. To show this, we do not work out a thermodynamic or quantum model of the laser ab initio, but exploit a number of mathematical similarities between the formulas expressing the maximum power generated by an endoreversible engine on one side and by a laser on the other side. No quantum model is involved or required. This strategy turns out to be fruitful, since it leads to meaningful considerations about the energy and entropy balance. We also will allocate *positive* absolute temperatures to the different stages of the machine, but these temperatures are not reservoir temperatures. We will find back using our method that at the laser threshold, no output power is generated, but that the machine there works at maximum efficiency. Using this new methodology, the laser model can be made accessible to thermodynamic arguments.

## 2. The endoreversible heat engine

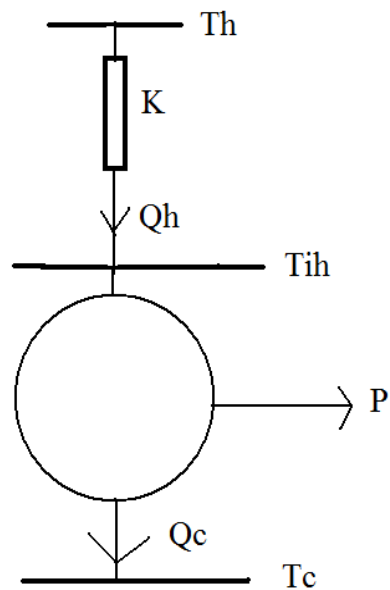

First, let us summarize the analysis of an endoreversible machine in its most simple form. The engine is connected to a hot reservoir at temperature $T_H$, which transfers power to the machine by e.g. a heat exchanger. At its other side, the heat exchanger is in contact with the (reversible) machine at temperature $T_{iH}$. This heat exchanger is responsible for the energy and entropy flow from the high temperature reservoir to the machine. It does this in a finite time, contrary to the infinite time it takes to transfer energy in the isothermal stages of a Carnot engine. The machine is cooled by a reservoir at temperature $T_C$, which is in direct contact (so no intermediary heat exchanger is involved) with the machine. This is a deliberate simplification of reality, since also at the low temperature side of the machine, a temperature gradient is needed to transport energy and entropy. But it makes life easier to go through the required algebraic analysis. The heat entering the machine through the heat exchanger at the hot side is given by the classical heat exchanger formula

$$q_H = K(T_H - T_{iH}) \tag{3}$$

where K summarizes the heat transfer properties of the heat exchanger and is expressed in W/K. The power generated by the engine is

$$P = q_H - q_C \tag{4}$$

where $q_C$ is the heat rejected into the cold reservoir. The efficiency $\eta$ of the machine is defined by how much incoming heat has been converted to power:

$$\eta = \frac{P}{q_H} = 1 - \frac{q_C}{q_H} \tag{5}$$

The machine itself is supposed to be working reversibly, no entropy is generated inside it so that the entropy influx to the machine equals the entropy outflux from the machine:

$$\frac{q_H}{T_{iH}} = \frac{q_C}{T_C} \tag{6}$$

Substituting Eq.(6) into eq.(4), we arrive at

$$P = q_H(1 - \frac{T_C}{T_{iH}}) = K(T_H - T_{iH})(1 - \frac{T_C}{T_{iH}}) \tag{7}$$

which is composed of three factors. Working out the two last ones, we find

$$P = K(T_H + T_C - \frac{T_H T_C}{T_{iH}} - T_{iH}) \tag{8}$$

Substituting Eq.(8) in the expression for the machine efficiency, Eq.(5), we find:

$$\eta = 1 - \frac{T_C}{T_{iH}} \tag{9}$$

By setting

$$\frac{dP}{dT_{iH}} = 0$$

we look for the extrema of the function (8). This turns out after a few algebraic manipulations, to result into a single maximum at the temperature

$$T_{iH} = \sqrt{T_H T_C} \triangleq T_{opt} \tag{9a}$$

$T_{iH}$ is called the control parameter in finite-time thermodynamics. Substituting Eq.(9) back into Eq.(8) for the power of the machine, we arrive at the maximum power generated by the endoreversible machine:

$$P_{max} = K(\sqrt{T_H} - \sqrt{T_C})^2 \qquad (10)$$

The engine efficiency at this power maximum is given by Eq.(9,10) as:

$$\eta_{CA} = 1 - \sqrt{\frac{T_C}{T_H}} \qquad (11)$$

and is called the Curzon-Ahlborn efficiency.

If the machine had been working completely reversibly (so that $T_H = T_{iH}$), then the entropy flux had been

$$\frac{q_H}{T_H} = \frac{q_C}{T_C} \qquad (12)$$

instead of eq.(6), and the efficiency would had been given now by the Carnot efficiency

$$\eta_{Carnot} = 1 - \frac{T_C}{T_H}. \qquad (13)$$

Note that the Curzon-Ahlborn efficiency at maximum power (11) is *lower* than the Carnot efficiency (13).

### 3. The three-level laser

We consider a three-level laser system, where the terminal laser level is the ground level $E_1$ of the active element. The laser is pumped to the pump level $E_3$ of the active medium and from there the pump level relaxes down nonradiatively to the upper laser level $E_2$ on a time scale which is small versus the radiative decay time of the upper laser level (i.e. the spontaneous lifetime of the upper level). This condition is fulfilled if the upper laser level is a metastable level of the dopant. In symbols:

$$\tau_{32} \ll \tau_{21}$$

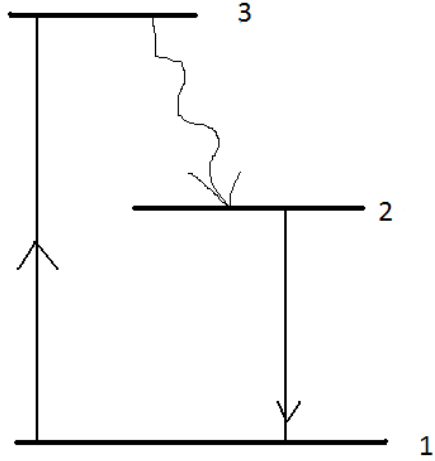

The energy released during the 3->2 transition is released as heat to the host material (or in quantum terms: this energy is creating phonons). Trough the pumping process, level 2 is gradually filled above its thermal equilibrium value, and a population inversion between level 2 and 1 develops. We do not take into account the pumping efficiency of the laser, i.e. how much of the incident pump power (be it electric or optic or else) is converted to power stored in the population inversion $P_{inv}$ inside the active medium, since this is a pure engineering issue without further physical relevance for our analysis. The cross section A of the lasing mode is supposed to be constant along the length d of the active medium in the resonator. The laser line itself is homogeneously broadened. The power accumulated in the inversion of the dopant at the laser threshold is noted as $P_{inv,th}$. The expressions for $P_{inv}$ and $P_{inv,th}$ are given by [4,5]

$$P_{inv} = AI_s g_0 d \qquad (14)$$

$$P_{inv,th} = -AI_s \ln(V\sqrt{R}) \qquad (15)$$

where $I_s$ is the saturation intensity (in W/m$^2$), $g_0$ is the unsaturated gain coefficient (in m$^{-1}$), V are the cavity losses and R is the reflection coefficient of the resonator output coupler. The output power of the laser is given by

$$P = -AI_s \ln(\sqrt{R})(\frac{P_{inv}}{P_{inv,th}} - 1) \qquad (16)$$

or, taking into account eq.(14) and eq.(15),

$$P = AI_s \ln(\sqrt{R})(\frac{g_0 d}{\ln(V\sqrt{R})} + 1) \tag{17}$$

These equations have been derived by adopting the (valid) supposition and justified simplification that above the laser threshold, the spontaneous decay from the upper level 2 to the lower level 1 becomes negligible compared to the stimulated decay. In other words, this stage is reversible, as required by the endoreversible machine model. In fact, we also have to suppose that the radiation entropy of the laser radiation is zero. This is again fulfilled to a high degree of accuracy, due to the high directivity (i.e. low divergence) of the monochromatic laser beam. In other words, in the spatial and temporal domain, the laser shows a nearly delta-function behavior. In thermodynamic terminology, laser radiation is pure work since it carries no entropy.

If the cavity losses V are now written in their exponential form as

$$V = \exp(-\alpha d)$$

then the output power is given as a function of the cavity parameters as:

$$P = AI_s \ln(\sqrt{R}) \left[ \frac{g_0 d}{\ln(\sqrt{R}) - \alpha d} + 1 \right] \tag{18}$$

The output power of the laser can now be optimized (maximized) as a function of the reflectivity R of the output coupler. This is done by taking the derivative with respect to R of eq.(18), which results in:

$$P_{max} = AI_s d(\sqrt{\alpha} - \sqrt{g_0})^2 \tag{19}$$

Note that this expression is invariant under a permutation of $g_0$ and $\alpha$.

The value of the optimal reflectivity is given by

$$\ln(\sqrt{R_{op}}) = d\sqrt{\alpha}(\sqrt{\alpha} - \sqrt{g_0}) \tag{20}$$

Eq.(19) and (20) are well-known relations in the laser literature.

### 4. The three-level laser as an endoreversible engine

Eq.(10) for the optimum machine power and eq.(19) for the optimum laser power are identical in mathematical form. It is not immediately obvious though how the control parameter $T_{iH}$ of the endoreversible engine is linked with the reflectivity R of the laser. So how can we find this link? Is it possible to start from eq.(10) and work backwards to restitute a resonator power output equivalent to eq. (19)? This would require the formal substitutions

$$\begin{aligned} AI_s d &\rightarrow K & AI_s d &\rightarrow K \\ g_0 &\rightarrow T_H \quad \text{or alternatively:} & g_0 &\rightarrow T_C \\ \alpha &\rightarrow T_C & \alpha &\rightarrow T_H \end{aligned} \quad (21)$$

We now define the control parameter for the laser as

$$x = -\frac{\ln(\sqrt{R})}{d} + \alpha \quad (22)$$

Physically, x is a measure of the total passive losses of the cavity. It is the sum of the output coupling "losses" (seen from the point of view of the laser, the power which is coupled out is an energetic loss) plus the other passive losses such as absorption, scattering and diffraction. The identification of this expression as control parameter, turns out to be crucial to unveil the three-level laser as an endoreversible machine. We re-formulate the output power Eq.(18) by using this parameter

$$\begin{aligned} P &= AI_s \ln(\sqrt{R})(\frac{-g_0}{x} + 1) \\ &= AI_s d(\alpha - x)(1 - \frac{g_0}{x}) \end{aligned} \quad (23)$$

This last formula would indicate that the formal allocation (21a) would be wrong since if we compare now the last two factors of Eq.(23) with those of Eq.(7), then apparently $T_H$ would correspond to $\alpha$ and $T_C$ to $g_0$. However, Eq.(23) is permutation-invariant: it can be rewritten without affecting the value of P, by interchanging $\alpha$ and $g_0$, which results in

$$P = AI_s d(g_0 - x)(1 - \frac{\alpha}{x}) \quad (24)$$

Eq.(24) now matches exactly in form with Eq.(7) and hence the left-hand side formal substitution (21) is the correct one. Note that Eq.(19) is invariant under a permutation of the parameters $\alpha$ and $g_0$, just as Eq.(24). The expression for the optimum control parameter is given by

$$x_{op} = \sqrt{\alpha g_0} \quad (25)$$

which is again invariant under a permutation of the parameters $\alpha$ and $g_0$ and formally similar to Eq.(9).

Finally, the control parameter x is frequently noted in the laser literature as

$$x = \frac{1}{c\tau_c}$$

where $\tau_c$ is called the lifetime of the photons in the cavity.

## 5. Thermodynamic considerations: the first law

The endoreversible three-temperature engine equivalent to the three-level laser is defined as the machine having the same output power of the laser, and operating between the "temperatures" $g_0$ and $\alpha$ (i.e. temperatures expressed in cm$^{-1}$ instead of in K). This equivalence was established by defining an appropriate control parameter x which determines the same maximum power output for both devices.

The machine will only work if the temperature of the hot reservoir is higher than the temperature of the cold reservoir which is nothing else than the population inversion condition for the laser. It should be noted that in the case of a three-level laser, this condition is not sufficient to start laser action: the inversion induces optical gain, which must be sufficient to overcome the cavity losses. This happens when the laser threshold condition is fulfilled, where the gain equals the total losses:

$$g_0 = \alpha - \frac{\ln\sqrt{R}}{d} = x \qquad (26)$$

Reverting the opening argument of this paragraph, it should be concluded that an endoreversible machine has also a threshold. This is clear from the figures in [7].

## 6. The second law

The entropy creation rate $\dot{S}$, expressed in W/K, is defined [7] by

$$\dot{S} = \frac{Q_C}{T_C} - \frac{Q_H}{T_H} \qquad (27)$$

or as a function of the machine efficiency:

$$\dot{S} = K\frac{(T_H - T_C - T_H\eta)^2}{T_H T_C (1-\eta)} \qquad (28)$$

where the power generated by the machine is in function of $\eta$:

$$P = K\frac{\eta(T_H - T_C - T_H\eta)}{1-\eta} \qquad (29)$$

We see that when $\eta$ equals the Carnot value eq. (1), P and $\dot{S}$ are both zero. The first fact points to the appearance of the laser threshold. The second fact points to the reversible operation of the machine at the Carnot limit.

Hence, at the laser threshold, the laser works with Carnot efficiency, but produces no work.

The entropy creation rate at the Curzon-Ahlborn efficiency is:

$$\dot{S}_{opt} = K\frac{\left(\sqrt{T_H} - \sqrt{T_C}\right)^2}{\sqrt{T_H T_C}}$$

Or, taking into account eq.(9,10), this can compactly be rewritten as

$$\dot{S}_{opt} = \frac{P_{max}}{T_{opt}}$$

Further, it is easy to see that eq.(27) can be rewritten as

$$\dot{S} = Q_H(\frac{1}{T_{iH}} - \frac{1}{T_H})$$

which indeed shows that the irreversibilities are concentrated in the heat exchanger. Also, by using this expression to work out the quantity $P + \dot{S}T_C$, we find that

$$P + \dot{S}T_C = Q_H\eta_C \triangleq e \qquad (30)$$

So, e is the maximum available work which can be done by the machine, it is called the exergy. The quantity $\dot{S}T_C$ is hence the power lost due to the irreversibilities. Expression (30) is known as the Gouy-Stodola Theorem [18]. We can work it out further in a particularly clear alternative expression by dividing eq.(28) through eq.(29), and multiplying with $T_C$:

$$T_C \frac{\dot{S}}{P} = \frac{\eta_C - \eta}{\eta} \tag{31}$$

## 7. Extending the analogy between a heat engine and a laser

Until now, we only found a mathematical similarity in the formulas describing a 3-temperature endoreversible heat engine and a 3-level laser, where the "temperatures" are expressed in inverse centimeters. We will now try to make the picture more physical in nature by identifying the input and output powers of the endoreversible model with the input and output powers of the tree-level model according to the allocation:

$$Q_H = \text{input power to the reversible stage}$$
$$Q_C = \text{power not extracted as laser power} \tag{32}$$
$$W = \text{laser output}$$

With this allocation of powers, the first law is valid:

$$Q_H = W + Q_C$$

which is identical to the equation for the endoreversible machine. We will now allocate temperatures to the different stages/levels of the laser. These temperatures however are not linked to a thermodynamical scale, where the population densities of the different levels are defined through the Boltzmann distribution. We are looking for an endoreversible engine, defined by the three temperatures $T_H, T_{iH}, T_C$ where we put in an amount of $Q_H$ of power, which results in a work output of W. Reconsidering again eq.(21), the following definitions have the dimension of temperature:

$$T_H = g_0 \frac{AdI_s}{K}$$
$$T_{iH} = x \frac{AdI_s}{K} \tag{33}$$
$$T_C = \alpha \frac{AdI_s}{K}$$

Looking back to eqs.(23) and (28), we see that the three laser parameters $\alpha$, x and $g_0$ are depending on the four engine parameters $T_H$, $T_{iH}$, $T_C$ and K. So we can only fix the values of say $KT_H$, $KT_{iH}$ and $T_C/T_{iH}$. So there are an infinity of endoreversible machines equivalent to one laser. By arbitrarily allocating a fixed value to, say, $T_C$, we can fix the value of the three other engine parameters.

It is unsatisfying however, that there still is a degree of freedom in the choice of the temperatures, since we have arbitrarily fixed $T_C$. This can be worked away, by realising that we have more equations coming from laser theory to express this quantity. It is therefore necessary to study the rate equations of the three-level laser.

We now convert the power accumulated in the population inversion, given by eqs. (14,15) to the new temperature terminology:

$$P_{inv} = AI_s d g_0 = KT_H$$

and similarly

$$P_{inv,th} = -AI_s (\ln(\sqrt{R}) - \alpha d) = AI_s x d = KT_{iH}$$

Hence we see that the exact meaning of $Q_H$ is given by

$$Q_H = K(T_H - T_{iH}) = P_{inv} - P_{inv,th} \tag{34}$$

This relation forces us to reconsider our terminology as defined by eq. (27). $Q_H$ is the power with which the reversible stage of the endoreversible machine is pumped. But it is the input power $P_{inv}$ which is stored in the inversion of the laser. The laser engineer will hence rather call $P_{inv}$ the pump power of the laser. $Q_H$ might more appropriately defined as the excess power. Let us in a similar way look to the meaning of $Q_C$:

$$Q_C = \frac{T_C}{T_{iH}} Q_H = \frac{\alpha}{x} Q_H \tag{35}$$

$Q_C$ hence represents the losses excluding the output coupling losses. Or, stated alternatively, it is the portion of the available power which is going to the passive cavity losses, excluding the output coupling losses.

We now consider the slope efficiency of the laser. It is defined by

$$\zeta = \frac{P}{P_{inv} - P_{inv,th}}$$

By using Eq.(29), we obtain that

$$\zeta = \frac{P}{Q_H}$$

Hence, the slope efficiency of the laser is equal to the regular efficiency $\eta$ of its equivalent endoreversible heat machine. We can also look at the regular efficiency of the laser, defined as

$$\zeta' = \frac{P}{P_{inv}}$$

$$\zeta' = \left(1 - \frac{T_{iH}}{T_H}\right)\zeta$$

$$\zeta' = \eta_{CA}^2$$

How physical are these newly defined temperatures? Needs further investigation. In each case, they give the correct power output and the correct Carnot and Curzon-Ahlborn efficiencies, since these last two parameters depend on the ratio $T_C/T_H$

### 8. An alternative three-level laser and a four-level laser

It is known that a second type of three-temperature endoreversible engine can be considered, where the heat exchanger is not located at the high-temperature side, but at the low-temperature side. Its power output is, following exactly the same procedure as for the first type, now given by

$$P = K(T_{iC} - T_C)(\frac{T_H}{T_{iC}} - 1)$$

instead of eq. (12). $T_{iC}$ can be eliminated through the maximum power condition

$$\frac{dP}{dT_{iC}} = 0$$

resulting finally in the same expression as eq.(9) for the output of the engine.

It is equally known that a second type of three-level laser exists, which is directly pumped into its upper level, and where the lower level de-excites towards the ground level. *Inversion is realized by the fast emptying of the lower laser level.* Remember that in our initial three-level model considered until now, the pump fills a pump level, which then quickly de-excites to the upper laser level, and the lower laser level is equal to the ground level. So there, *inversion was realized by the fast filling of the upper laser level.*

So in fact it comes as no surprise that the three-temperature endoreversible machine where the heat exchanger is located at the low temperature side, is equivalent to the second type of three-level laser.

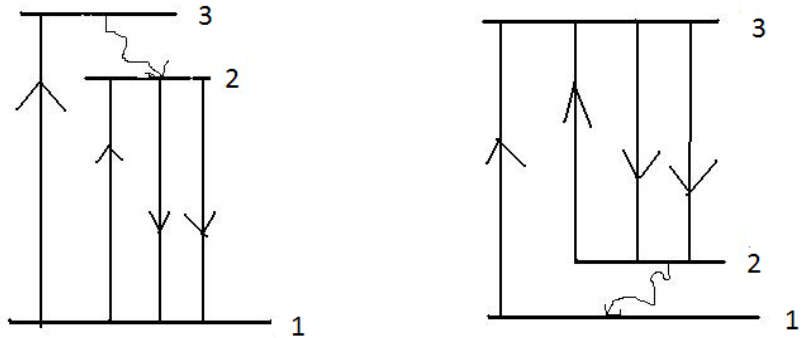

The expression eq.(17) specifying the output power of a three-level laser, is in fact universally valid, independent of the level structure. It can be derived for any kind of resonator containing an amplifying medium. So we can apply it as well to the four-level laser. It is evident that the endoreversible counterpart of a four-level laser (levels 1, 2, 3 and 4, lasing is occurring from 3 to 2) is a four-temperature machine ($T_H, T_{iH}, T_C, T_{iC}$) with a heat exchanger at the hot side, and a heat exchanger at the cold side.

Algebraically, this makes our life difficult, since we have now two control parameters, $T_{iC}$ and $T_{iH}$ which need to be optimized simultaneously, in order to find the maximum power. In the book of H. Fuchs " The dynamics of heat", an alternative way is followed, not by maximizing the power, but the problem is re-formulated as a minimization of the entropy currents:

$$\frac{d\Pi_S}{dQ_H} = 0$$

where $\Pi_S$ is the sum of the entropy currents going through the two heat exchangers, *plus an entropy current going directly from the hot to the cold reservoir.* The path of the energy current flows is hence not identical to the path of the entropy current flows. I have re-derived the equations for the case of a three-temperature machine ($T_H, T_{iH}, T_C$). This eliminates the entropy current through the low-temperature heat exchanger, and simplifies the mathematics, putting in a clearer light the point which Fuchs wants to make. The sum of the entropy current through the high-temperature heat exchanger plus the direct entropy current not going through the reversible part is hence:

$$\Pi_S = Q_H(\frac{1}{T_{iH}} - \frac{1}{T_H}) + (Q^* - Q_H)(\frac{1}{T_C} - \frac{1}{T_H})$$

where $Q^*$ is the total heat entering the machine and is supposed to be constant. We eliminate $T_{iH}$ from this equation, using the heat exchanger equation (3):

$$\Pi_S = Q^*(\frac{1}{T_C} - \frac{1}{T_H}) + Q_H\left(\frac{1}{T_H - \frac{Q_H}{K}} - \frac{1}{T_C}\right)$$

Taking the derivative to the control parameter $Q_H$ in this equation, and setting it to zero, we arrive at a quadratic equation for $Q_H$:

$$Q_H^2 - 2KT_HQ_H + K^2(T_H^2 - T_HT_C) = 0$$

Here, we choose the root with the negative sign so that $Q_H$ is cast in the form of the heat exchanger equation, we obtain:

$$Q_H = K(T_H - \sqrt{T_HT_C})$$

This gives for the work done by the machine:

$$P = Q_H - Q_C$$
$$= Q_H - \frac{T_C}{T_{iH}}Q_H$$

by the endoreversibility of the machine, leading to the optimal power output of

$$P = K(\sqrt{T_H} - \sqrt{T_C})^2$$

which is identical to eq. (10). In this way, also the Curzon-Ahlborn machine efficiency is found back.

Fuchs remarks that A. De Vos has found in 1985 a simple derivation of the CA expression for the four-temperature endoreversible machine, based on power maximalization instead of entropy minimalization. For the case of the laser, we might be tempted to choose two different heat exchangers, but if we look to eq.(21), we see that the heat transfer coefficient only depends on the volume of the host and the saturation intensity of the dopant. We are hence forced to fix the same heat transfer coefficient for both heat exchangers. This is compatible with the fact that the heat generated in the transition 4->3 and the transition 2->1 is taking place in the same host material. Further, we have only three laser parameters, α, x and $g_0$ to define four temperatures. But we should not forget that $T_{iC}$ is not independent of $T_{iH}$, because they are connected through the iso-entropy condition. We take the expression over from ref.[8]:

$$T_{iC} = \frac{T_C \cdot T_{iH}}{2T_{iH} - T_H}$$

Apparently, this parameter does not have a direct physical interpretation in terms of the laser parameters. What it does demonstrate however is that, if the laser transition is working reversibly, the two heat inputs to the host material are interconnected through a thermodynamical condition.